\documentclass[a4paper,12pt]{article}
\usepackage[utf8]{inputenc}
\usepackage{amsmath, amssymb}
\newcommand{\Hca}{\mathcal{H}}

\title{Radiative Corrections as Origin of Tiny Fermion Masses}
\author{A K Kapoor\footnote{akkhcu@gmail.com}\\
296 Doyen's Colony\\
SeriLingampally, Hyderabad 500019, INDIA}

\begin{document}

\maketitle

\baselineskip=18pt

\begin{abstract}
The fermion masses in the standard model are introduced as arbitrary parameters
and there is no understanding of their origin. In this letter it is suggested
that small non zero neutrino masses may be a reflection  of broken stochastic
supersymmetry that guarantees  the equivalence of Parisi Wu stochastic
quantization scheme to standard quantum field theory.
\end{abstract}

All predictions of the standard model of electroweak unification \cite{SM,
Ellis} have been
beautifully confirmed by experiments so far. With the discovery of Higgs boson,
the last missing link has been found. For a long time, until the discovery  of
neutrino oscillations, it had been widely believed that the neutrinos are
massless and are adequately described by two component Weyl spinors.
Renormalizability of the theory has been ensured by  presence of axial
anomaly cancellation mechanism between  different fermion sectors.
Subsequent confirmation of  neutrino oscillations have changed this scenario
completely. The
observation of neutrino oscillations is universally seen as a signal
of new physics beyond standard model. For a review and
references on models beyond the standard model see references
\cite{ Ellis,RNM,Parker, Elor}.
Going beyond the standard
model, one may have to introduce either Majorana neutrinos or four component
Dirac neutrinos. In case of Dirac neutrinos  right handed components
of neutrinos must be included in the theory. In either case, the
neutrino masses are not constrained by the
symmetries of the model and the  masses have remained free
parameters.

Theoretical  models have been suggested to
explain why neutrino masses could be small.
Possible theoretical explanations,  such as see-saw mechanism
introduce new particles which will be seen at a higher energy
scale. Review of models proposed and references can be found in \cite{King1,
King2}. These
models do not offer any insight into the fermion mass problem and do not
predict the neutrino masses, and do  not offer any clues as to
why neutrino masses are so tiny. Most of the present efforts to understand the
neutrino masses are centred around determining the masses from the
experimental data.

In this article we suggest a possible mechanism for explanation of origin of
tiny neutrino masses staying as close to the standard electroweak model as
possible.
It is suggested that the neutrino masses at the tree level are zero and non
zero  masses arise purely out  of radiative corrections. Of course having
non-zero mass requires a four component Dirac neutrino. As already remarked
this suggestion goes beyond the standard model. Whether  a model, with
Dirac neutrinos, eventually explains the  masses can be decided only by
building a detailed model and comparing detailed theoretical predictions with
experiments.

Instead of going beyond the standard model and looking for new interactions
responsible for neutrino masses, it is suggested to
look for quantization schemes beyond the standard quantization
scheme, formally equivalent to canonical quantization. This scheme should have
features that it
should in some limit reproduce all the known results and should offer a
possibility of tiny neutrino masses arising out of radiative
corrections. Such an approach will predict the masses without any free
parameters and can be tested against the experiments.

 The conventional quantum field theory
(CQFT) formalism, formulated using canonical quantization,
treats a massless fermion with four components as if the theory has a left
handed and a right handed massless fermion  and a consistent formalism is also
possible only with two components. This is
primary the reason that   the CQFT is equally well equipped to describe  two
component Weyl, Majorana neutrino, or a four component Dirac neutrino. In
absence of any other restriction in the standard model, this means that
the fermion masses become free parameters. In this letter it is suggested that
this situation will change when we go to Parisi Wu stochastic
quantization method(SQM) \cite{SQM1, SQM2,SQM3, SQM4}. Though it is widely
believed that SQM is equivalent
to the CQFT, this equivalence has been demonstrated mostly at a formal level.

For  models with bosons only stochastic supersymmetry, found in  Parisi Wu
formulation of SQM, plays a crucial role in proof of  equivalence of SQM and
conventional formulations \cite{Nakaz,VSAKSC}. In absence of stochastic
supersymmetry the equivalence proofs will not be applicable. The initial
investigations in SQM have been driven by the fact that SQM offered an
scheme for gauge theories without gauge fixing. Though the presence  of
Zwanziger gauge fixing, suggested  later, does not respect the
stochastic supersymmetry, the equivalence of pure Yang Mills in SQM has been
investigated, and a proof of equivalence of gauge
invariant amplitudes has been given in \cite{Zwan} and  renormalization
of pure Yang Mills theory has been studied in \cite{Zinn}. However,
a detailed investigation and a proof of equivalence  of gauge
theories including fermions is not available.

The earliest formulation of fermionic
theories in SQM was given in \cite{Fukai}. {\it It turns out
the SQM of gauge theories in presence of fermions is not  always
equivalent to CQFT.}
In context of electroweak interactions, a proof of equivalence of SQM and CQFT
at  perturbative level has not been given. A physical field theory is
completely defined by the Lagrangian, a way of handling divergences and by
rules used to extract
finite answers. Even though the equivalence between Parisi Wu SQM and
conventional formulations of quantum field theory has been a subject of intense
investigation the actual situation could turn out to be different from expected
behaviour for reasons outlined below.

A closer look at the fermionic theories several reveals
differences in the SQM and CQFT formalism. To begin with, a consistent
formulation of SQM for fermions appears to require introduction of all
components.
A consistent SQM formulation of four component  Dirac fermion
always violates
chiral symmetry even at tree level. Also  it has
almost gone unnoticed that a  renormalized theory of Dirac fermions based on SQM
\cite{Fermi1,Fermi2}
has some features very different from renormalized theory based on CQFT.
In particular, new counter terms  may be
required which destroy the stochastic supersymmetry through the radiative
corrections.

The three  features of the SQM formalism,
requiring use of four component Dirac fermions, violation of  chiral
symmetry even at tree level, and the renormalized theory not
preserving stochastic supersymmetry will be demonstrated
taking example of Yukawa scalar coupling and of a model with axial
vector coupling respectively.

The main features of SQM formulation will  be first summarized.  The
CQFT of a fermion is described by the Lagrangian
\begin{eqnarray}
    L&=&  \bar{\psi}(i\gamma_\mu \partial_\mu + M ) \psi + L^\prime
\end{eqnarray}
where $L^\prime$ is part of the Lagrangian describing other fields which may be
coupled to the fermion. So for a scalar field with Yukawa coupling
$L^\prime=L_1$ where
\begin{eqnarray}
    L_1 &=& \frac{1}{2} [ (\partial_\mu \phi)(\partial_\mu \phi) + m^2 \phi^2]
    + \lambda \bar{\psi}\psi \phi
     +\frac{g}{4!}\phi^4\label{EQ01}.
\end{eqnarray}
The SQM formulation makes use of Euclidean action $S$ corresponding to the
Lagrangian $L$. The SQM formulation of scalar field theory with Yukawa coupling
and renormalization has been discussed in detail in \cite{Phi4}. Here we will
recall basic equations and important features only and for details we refer to
the original articles. The basic Langevin equations of SQM are given by
\begin{eqnarray}
  \frac{\partial \phi(x,t)}{\partial t}
  &=& - \gamma^{-1} \frac{\delta S}{\delta \phi} + \eta(x,t)\label{EQ02}\\
 \frac{\partial \psi(x,t)}{\partial t}
 &=& - \int dx^\prime K(x,x^\prime) \frac{\delta S}{\delta
\bar{\psi}(x^\prime)} + \theta(x,t)\label{EQ03} \\
\frac{\partial \bar{\psi}(x,t)}{\partial t}
 &=&  \int dx^\prime  \frac{\delta S}{\delta \bar{\psi}(x^\prime)} K(x,x^\prime)
+ \bar{\theta}(x,t)\label{EQ04}.
\end{eqnarray}
Here $x$ collectively stands for all the components of the Euclidean four
vector $x_\mu$ and
$t$ denotes the fifth time or the stochastic time. The Gaussian white noises
$\eta(x,t), \theta(x,t), \bar{\theta}(x,t)$ are assumed to have averages
\begin{eqnarray}
  \langle\eta(x,t)\eta(x^\prime,t^\prime) \rangle =
2\gamma^{-1}\delta(x-x^\prime)\delta(t-t^\prime) \label{EQ05}\\
  \langle\theta(x,t)\bar{\theta}(x^\prime,t^\prime) \rangle =
2 K(x,x^\prime)\delta(t-t^\prime) \label{EQ06}
\end{eqnarray}
In the operator formalism \cite{NMYY} of SQM, this theory is equivalent to
a five dimensional field theory described by a five dimensional stochastic
action given by
\begin{equation}
 \Lambda = \int dx dt\left( \pi \frac{\partial\phi}{\partial t}
  + \frac{\partial\bar{\psi}}{\partial t}\omega  + \bar{\omega}
\frac{\partial\psi}{\partial t} - \Hca
 \right) \label{EQ07}
\end{equation}
where
\begin{eqnarray}
{\mathcal H}&=&T_\psi \left[ 2\bar{\omega}K \omega
-\bar{\omega}\tilde{K}\frac{\delta
S}{\delta \bar{\psi} } + \frac{\delta S}{\delta \psi}\tilde{K} \omega\right] +
\Hca_1 \\
\Hca_1 &=& \gamma^{-1}\left[\pi^2 - \pi(-\Box + m^2) \phi - \lambda \pi
\bar{\psi}\psi - \frac{g}{3!} \pi \phi^3  \right]
\label{EQ08}.
\end{eqnarray}
and $\pi, \omega$ are the stochastic momentum fields conjugate to
the scalar field $\phi$ and the fermion field $\psi$.

A simple, but important feature of the stochastic theory, that the above
equations bring out, is that, for every choice of a kernel $K$,  different
terms involving the fermion,
$\frac{\partial\bar{\psi}}{\partial t}\omega + \bar{\omega}
\frac{\partial\psi}{\partial t}$,   $ 2\bar{\omega}K \omega$
and $\bar{\omega}\tilde{K}\frac{\delta
S}{\delta \bar{\psi} } + \frac{\delta S}{\delta \psi}\tilde{K} \omega$
have different behaviour under axial transformations. This in turn
means a necessary mix up  the left and right components,  independent of the
interaction  Lagrangian chosen in four dimensional CQFT.  At the tree level the
stochastic action $\Lambda$ is not invariant under axial transformations even if
the underlying CQFT preserves the chiral symmetry.

The SQM cannot be formulated for a two component fermions because with two
component fermion, the Langevin equations will be inconsistent with the
requirement that the kernel $K(x, x^\prime)$ be invertible.

For the scalar Yukawa coupling, the  structure  of
counter terms and the renormalized  theory described by the five dimensional
action $\Lambda$ has been discussed in detail in \cite{Phi4}. In this case it
was shown that the stochastic supersymmetry, crucial to equivalence with CQFT,
 can be maintained. However it turns out that this is not always the case
for fermions \cite{Fermi1, Fermi2}.  A  concrete support for this statement can
be seen by taking a
specific example. Let us consider the case of and axial vector
$A_\mu$ coupled to the fermion. In this case $L^\prime$ would be equal to
$L_2$ where
\begin{equation}
  L_2 =-\frac{1}{4}F_{\mu\nu} F^{\mu\nu} - \lambda \bar{\psi}i\gamma^\mu
\gamma_5
           \psi\, A_\mu
\end{equation}
A straight
forward exercise in  power counting
reveals that the stochastic theory requires finite number of counter
terms.  Due to  divergent triangle  diagram, a new counter
term of the form $\epsilon_{\mu\nu\lambda\sigma}\pi_\mu A_\nu\partial_\lambda
A_\sigma$ is needed and this term preserves axial gauge invariance but
violates stochastic suspersymmetry. The appearance of this counter term is
similar to appearance of $\phi^4$ term in CQFT of scalar field $\phi$ coupled
with a fermion, in absence of a scalar self interactions at the classical level.

.

Even though the underlying the classical Lagrangian of a model may be the same,
an anomaly free CQFT for  vector and axial vector gauge coupling with four
component massless fermions is likely to be very different from the
renormalized theory within SQM framework.  It is possible to preserve zero mass
for a fermion in an anomaly free gauge theory with vector and axial vector
couplings, but SQM of the same model, even at the tree level, does not preserve
axial symmetry for massless fermions and does not guarantee a theory equivalent
to that  CQFT.

Barring some accidental cancellations, absence of both axial symmetry and
stochastic supersymmetry in SQM  will mean non zero radiative corrections to the
fermion masses and also calculable effects beyond those present the suitably
extended standard model in CQFT formalism.

The aim of this paper has been only to point out the possibility that the
neutrino masses could come purely out of radiative corrections. In order
to realize the mechanism suggested here a possible strategy will be as follows.
We first start by adding right handed
neutrinos, and ensure  a renormalizable CQFT with zero mass neutrino at tree
level by symmetry considerations with chiral symmetry ensuring zero masses for
the neutrinos at all orders. This will require inclusion of some new
features in the standard model. Having done this, SQM of such a model breaks
the chiral symmetry at the tree level  itself, and the stochastic supersymmetry
in higher orders. Thus using SQM, while the theory  remains
renormalizable, it will give different predictions. As the equivalence of CQFT
and SQM is lost in higher orders, the deviations in SQM from CQFT predictions
will come from radiative corrections and will be calculable.

Whether nature chooses this scheme or something else can be determined  solely
by a detailed computation and its comparison with the experiments.

A realistic model is most likely be complicated to analyse and is beyond the
scope of short communication like present letter and a complete analysis needs
to be taken up separately.

I thank Bindu Bambah for an illuminating discussion of present status of
research in neutrino physics which prompted this work and to H.S. Mani for
critical comments  on an
earlier draft of the manuscript.

\end{document}